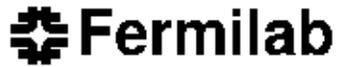 FERMILAB-Conf-04/056-AD   April 2004# ON THE THEORY AND SIMULATION OF MULTIPLE COULOMB SCATTERING OF HEAVY CHARGED PARTICLES[*]

S.I. Striganov

*Fermi National Accelerator Laboratory, MS 220, Batavia, Illinois 60510-0500, USA*April 29, 2004

## Abstract

The Moliere theory of multiple Coulomb scattering is modified to take into account difference between scattering off atomic nuclei and electron. A simple analytical expression for angular distribution of charged particles passing through a thick absorber is found. It does not assume any special form for a differential scattering cross section and has wider range of applicability than a Gaussian approximation. A well-known method to simulate multiple Coulomb scattering is based on the different treatment of "soft" and "hard" collisions. An angular deflection in a large number of "soft" collisions is sampled using the proposed distribution function, a small number of "hard" collision are simulated directly. A boundary between "hard" and "soft" collisions is defined providing a precise sampling of a scattering angle (1% level) and a small number of "hard" collisions. A corresponding simulating module takes into account projectile and nucleus charged distributions and exact kinematics of a projectile-electron interactions.---

[*]Presented paper at the *10th International Conference on Radiation Shielding*, Funchal (Madeira), Portugal, May 9-14, 2004.

# ON THE THEORY AND SIMULATION OF MULTIPLE COULOMB SCATTERING OF HEAVY CHARGED PARTICLES

S.I. Striganov

*Accelerator Division, Fermi National Accelerator Laboratory, MS 220, Batavia, Illinois 60510-0500, USA*
Phone +1-630-840-2374, fax +1-630-840-6039, e-mail: strigano@fnal.gov

**Abstract** – The Moliere theory of multiple Coulomb scattering is modified to take into account difference between scattering off atomic nuclei and electron. A simple analytical expression for angular distribution of charged particles passing through a thick absorber is found. It does not assume any special form for a differential scattering cross section and has wider range of applicability than a Gaussian approximation. A well-known method to simulate multiple Coulomb scattering is based on the different treatment of "soft" and "hard" collisions. An angular deflection in a large number of "soft" collisions is sampled using the proposed distribution function, a small number of "hard" collision are simulated directly. A boundary between "hard" and "soft" collisions is defined providing a precise sampling of a scattering angle (1% level) and a small number of "hard" collisions. A corresponding simulating module takes into account projectile and nucleus charged distributions and exact kinematics of a projectile-electron interactions.

## INTRODUCTION

Multiple scattering of charged particles in the Coulomb field of nuclei is of interest for numerous applications related to particle transport in matter. A comprehensive comparison of the Moliere theory[1] with experimental data on multiple scattering of 1 MeV to 200 GeV protons shows that this theory, with the Fano correction[2], is accurate to better than 1% on average[3] except for thick absorbers[4] and for hydrogen targets at high energies[3].

## NEW ASYMPTOTIC FOR THICK SCATTERERS

An angular distribution of a charged particle after passing through an absorber of a length $t$, can be written as

$$F(\theta,t) = \frac{1}{2\pi}\int_0^\infty J_0(p\theta)\exp(-tA(p))p\,dp, \quad (1)$$

where

$$tA(p) = t\int_0^\infty d\Omega(1 - J_0(p\theta))\frac{d\Sigma}{d\Omega}, \quad (2)$$

here $\frac{d\Sigma}{d\Omega}$ is a single scattering cross section. If thickness $t$ is large enough, then only small $p$ are important in (1) and (2), then

$$tA(p) = \frac{p^2}{4}\langle\theta^2\rangle - \frac{p^4}{64}\langle\theta^4\rangle,$$

$$\langle\theta^k\rangle = t\int_0^\infty d\Omega\,\theta^k\,\frac{d\Sigma}{d\Omega}. \quad (3)$$

In this case, the angular distribution can be rewritten in a simple form

$$F(\theta,t) = \frac{\exp(-\phi)}{\pi\langle\theta^2\rangle}(1 + rL_2(\phi) + 3r^2 L_4(\phi) + \ldots), \quad (4)$$

where $L_k$ are the Laguerre polynomials.

$$L_2(x) = 1 - 2x + \frac{x^2}{2},\; \phi = \frac{\theta^2}{\langle\theta^2\rangle},\; r = \frac{\langle\theta^4\rangle}{2\langle\theta^2\rangle^2} \propto \frac{1}{t}.$$

For a large thickness, the parameter $r$ becomes small and one or two terms in (3) are enough to provide a reasonable accuracy.

Note, that the Moliere theory is based on a single scattering cross section, which has infinite moments (3). Therefore, this theory has a wrong asymptotic at large thicknesses. This problem can be overcome if a finite nuclear size is taken into account. The solution for an angular distribution was obtained[4] under an assumption that the charge distribution in nuclei is Gaussian. It coincides with Moliere for a thickness of 0.1-1 radiation lengths and reaches Gaussian for 100-1000 radiation lengths. This modification of the Moliere theory can be used to find a range of the applicability of the approximation (4). It is shown in Fig.1 that two first terms in (4) provide about 1% agreement with a precise calculation for four orders of magnitude at $r \geq 0.01$. One term in (4) (Gaussian distribution) is enough to describe first two decades only. For $r>0.05$, the new approximation with first two terms provides better than 10% accuracy. In more convenient units, range of the validity of this approximation can be estimated from

$$r \approx \frac{368 X_0}{(Z+6.4)t},$$

where $X_0$ is a radiation length and $Z$ is a charge of the absorber nuclei.

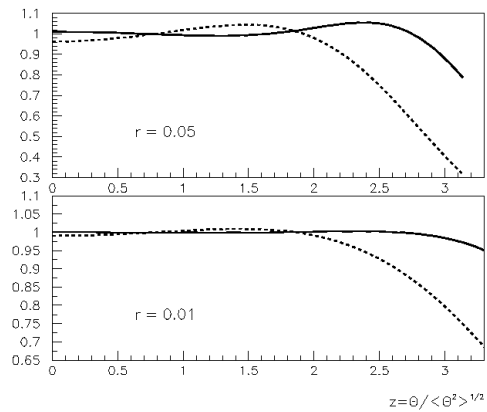

Figure 1. Ratio of asymptotic (4) and precise angular distribution[4]. Dashed line is a first term in (4), solid line is first two terms in (4).

## SCATTERING OFF ATOMIC ELECTRONS AND MOLIERE THEORY

A charged particle traversing medium is deflected primarily by elastic collisions in the Coulomb field of nuclei. Inelastic collisions with atomic electrons also should be taken into account. To estimate a contribution of inelastic collisions, Bethe proposed[1] to replace the squared nuclear charge $Z^2$ with the sum of the squares of the nuclear and electronic charges $Z(Z+1)$. This procedure would be accurate if the single scattering cross sections were the same for nucleus and electron targets. The actual cross sections are different at small and large angles. Let's consider a modification to the Moliere theory, which takes into account these differences.

Elastic scattering cross section reads[1]

$$t \frac{d\Sigma_{el}}{d\Omega} = \frac{\chi_c^2 q_{el}(\chi)}{\pi \chi^4},  \qquad (5)$$

$$\chi_c^2 = \frac{4\pi N t z^2 Z(Z+1) e^4}{A p^2 \beta^2}, \qquad (6)$$

where $ze, p, \beta$ are charge, momentum and velocity of the incident particle, $q_{el}(\chi)$ is a screening function, $N$ is the Avogadro's number and $A$ is an atomic weight of target material.

The recoil imparted to atomic electron by incident heavy particle cannot exceed a certain limit, so a simple approximation of the inelastic cross section is given by

$$t \frac{d\Sigma_{in}}{d\Omega} = \frac{\chi_c^2 q_{in}(\chi)}{Z \pi \chi^4}, \quad \chi \leq \chi_{max}, \qquad (7)$$

$\chi_{max}$ is so chosen, that the mean-squared angle resulting from (7) is adjusted to the mean-squared angle calculated from the precise cross section.

Using (1), (2), (5), (7), the angular distribution can be written as

$$F(\theta, t) = \frac{1}{2\pi} \int_0^\infty J_0(\frac{u\theta}{\theta_M}) \exp(-\frac{u^2}{4} + U(u, B)) u \, du$$

$$U = \frac{u^2}{4B} \log(\frac{u^2}{4}) + \frac{2u^2}{(Z+1)B} \int_{wu}^\infty dx \frac{1-J_0(x)}{x^3} \qquad (8)$$

$$w = \frac{\chi_{max}}{\theta_M}, \quad \theta_M^2 = B \chi_c^2.$$

$B$ is defined from

$$B - \ln B = \ln \frac{\chi_c^2}{\chi_a^2} + 1 - 2C + \frac{1}{Z+1} \ln \frac{\chi_a^2}{\chi_i^2} \qquad (9)$$

At small thicknesses ($w>>1$), the distribution (8) is described by the Moliere function with parameters defined from (6) and (9). At large thicknesses ($w<<1$), (8) also coincides with Moliere, but parameters are calculated using

$$B - \ln B = \ln \frac{\chi_{cM}^2}{\chi_a^2} + 1 - 2C + \frac{1}{Z} \ln \frac{\chi_{max}^2}{\chi_i^2},$$

$$\chi_{cM}^2 = \chi_c^2 \frac{Z}{Z+1} \qquad (10)$$

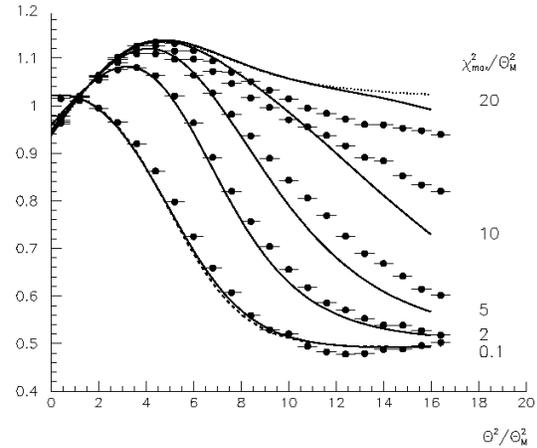

Figure 2. Ratio of the modified Moliere theory and Bethe approach for 10 GeV/c muon on a hydrogen absorber. Solid line is Eq.(8), dotted line is Eq.(9), dashed line is Eq.(10). Symbols are Monte Carlo simulations.

Fig.2 shows a ratio of angular distributions calculated using equation (8), (9), (10) and the Moliere distribution obtained using the Bethe $Z(Z+1)$ approach. It is seen that a *standard $Z(Z+1)$* approximation agrees well with more precise consideration (8) for small angles, but it overestimates angular distribution by 50% at large angles for hydrogen. Calculations for other targets show that this large-angle ratio is simply

$Z/(Z+1)$. The asymptotics (9) and (10) have been obtained by Fano[2]. He believed that solution (9) is valid for incident electrons and solution (10) can be applied for heavy particles. Our consideration shows that the above limits have different ranges of applicability. If $w \gg 1$, solution (9) should be used even for heavy particles. As shown in Table 1, this conclusion is supported by experiment.

Table 1. Ratio of measured and calculated widths of angular distributions.

| Z | w | Exp/Bethe | Exp/Fano(9) | Exp/Fano(10) |
|---|---|---|---|---|
| 1 | 9.7 | 0.99±0.01 | 0.96±0.01 | 0.88±0.01 |
| 4 | 10.3 | 1.02±0.04 | 1.00±0.04 | 0.96±0.04 |
| 6 | 9.8 | 1.03±0.04 | 1.01±0.04 | 0.98±0.04 |
| 4 | 0.12 | 1.00±0.06 | 0.98±0.06 | 1.03±0.06 |
| 6 | 0.10 | 0.97±0.04 | 0.95±0.04 | 0.98±0.04 |

Low-$w$ measurements[3] agree well with formula (10), but at large $w$ formula (9) is much closer to data[5] than the original Fano result (10). Note, that widths of angular distributions were measured at small angles ($\theta/\theta_M < 2$). Therefore, the Bethe approach looks like perfect in Table 1. But as was shown above, this approximation overestimates the large angle scattering.

## MONTE CARLO APPROACH

An efficient method to simulate multiple scattering is based on a separate treatment of "soft" and "hard" interactions. Angular deflection in a large number of "soft" collisions is sampled from a "continues" distribution, "hard" scatterings are simulated explicitly[6]. There is an obvious correlation between precision and efficiency of the algorithm and the value of a boundary angle $\theta_b$ between "soft" and "hard" collisions. For small $\theta_b$, a number of discrete interactions is large and precision is high, for larger $\theta_b$, the efficiency increases but the accuracy decreases. The optimal value of a boundary angle is important but still an open question.

It can be shown that the "continues" distribution is given by

$$F(\theta, t) = \frac{\exp(-\phi)}{\pi \langle \theta_s^2 \rangle}(1 + r_s L_2(\phi)), \quad (11)$$

where

$$\langle \theta_s^k \rangle = t \int_0^{\theta_b} d\Omega\, \theta^k \frac{d\Sigma}{d\Omega}; \quad \phi = \frac{\theta^2}{\langle \theta_s^k \rangle}; \quad r_s = \frac{\langle \theta_s^4 \rangle}{2\langle \theta_s^2 \rangle^2} \propto \frac{1}{t}.$$

A range of the applicability of (11) is defined as

$$r_s(\theta_b, t) = \delta \ll 1. \quad (12)$$

For any step of length $t$, the boundary angle $\theta_b$ can be calculated from equation (12). For elastic scattering using (5), (6), equation (12) can be rewritten as

$$x - \ln x = b = \frac{1}{2}(\ln\frac{8\delta\chi_{cM}^2}{\chi_a^2} + 1), \quad x = \frac{\theta_b^2}{8\delta\chi_{cM}^2}. \quad (13)$$

A solution of equation (13) with better than 10% accuracy is given by

$$x = \frac{b}{2}(1 + \frac{\ln b}{b-1})(1 + \sqrt{1 - \frac{1}{b}}). \quad (14)$$

The derivation between the exact and approximate solutions can be considered as a redefinition of $\delta$. For very small thicknesses (about 0.001 of a radiation length) equation (13) can not be resolved for an arbitrary $\delta$. In this case, a maximal possible value of $\delta$ can be used. To perform simulation with an arbitrary nuclear form factor, the boundary angle should be limited at a large step

$$\theta_b^2 \leq 0.1\theta_{nuc}^2 = \frac{0.3}{p^2 r_{nuc}^2}, \quad (15)$$

where $r_{nuc}^2$ is an average nuclear radius squared. At large thicknesses, a number of "hard" interactions becomes not small. In this case, the angular distribution is well described by the approximation (4). Scattering angles can be sampled directly from (4). A comparison of angular distributions simulated by the above algorithm and calculated using analytical solutions[1,4] shows that results agree within about 1% for $\delta = 0.03$. As shown in Fig.3, a number of "hard" elastic scatterings is small for this value of $\delta$.

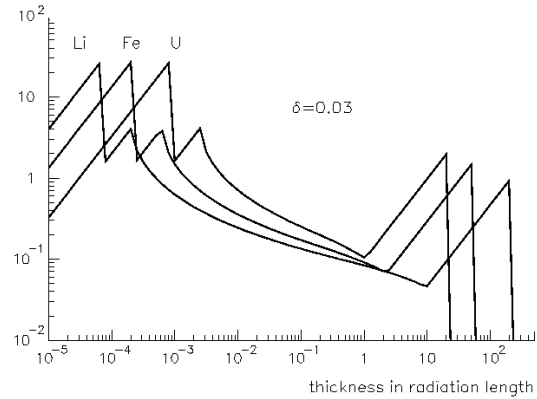

Figure 3. Average number of "hard" elastic collisions.

The boundary angle $\theta_{in}$ for inelastic collisions can be obtained also from (13). In this case

$$b_{in} = \frac{1}{2}(\ln\frac{8\delta\chi_{cM}^2}{Z\chi_i^2} + 1), \quad x = \frac{Z\theta_{in}^2}{8\delta_{in}\chi_{cM}^2} \quad (16)$$

The procedure of determination of the inelastic boundary angle was checked by comparison of Monte Carlo simulation based on simple cross sections (5), (6), (7) and analytical solution (8), (9). Results agree within about 1% if $\delta_{in} < 0.1$. Equations (13), (16) were obtained for small angles for which only the Ruzerford part of the cross section is important. So, the inelastic boundary angle should be limited by

$$\frac{\beta^2 \Delta_{\min}}{\varepsilon_{\max}} = \frac{\beta^2 p^2 \theta_{in}^2}{2 m_e \varepsilon_{\max}} = \gamma_{in} \ll 1, \qquad (17)$$

where $\Delta_{\min}$ and $\varepsilon_{\max}$ are the minimum and maximum energies transferred to electron in a single collision.

In inelastic scattering with electron, the projectile undergoes angular deflection *and* loses energy. For discrete inelastic collisions, the correlation between the energy loss and scattering angle is determined by kinematics. "Continuous" energy losses are described well by the Vavilov distribution with redefined parameters[7]. It is known[7] that a log-normal distribution fits well the Vavilov function for

$$\kappa = \frac{\xi}{\Delta_{\min}} = \frac{\chi_{cM}^2}{Z \theta_{in}^2} \geq 0.3 \cdot \qquad (18)$$

With the boundary inelastic angle determined from (16)-(18), the average number of inelastic discrete collisions is small, in most cases it is about $\kappa$. A comparison of simulations and the modified Moliere theory for hydrogen is shown in Fig.2. The theory based on the approximation (7) reproduces main features, but at large angles and small thicknesses Monte Carlo, which uses a more precise cross section, should be applied. Note, that in such an approach, the energy loss and angular distributions are correlated, at least for low-*Z* targets.

A comparison of the Monte Carlo simulation and analytical calculation for high-*Z* target is presented in

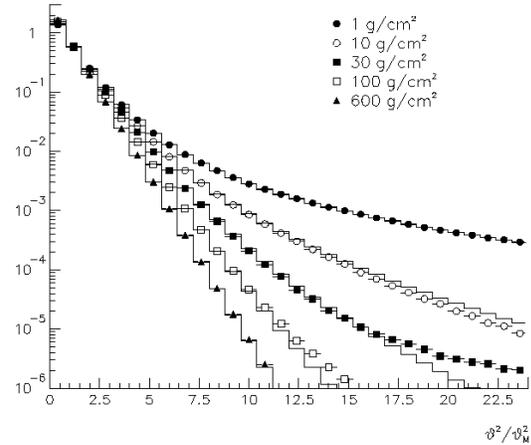

Figure 4. Angular distributions of 50 GeV/c muon passing through a uranium absorber. Symbols - Monte Carlo simulation, solid line – theory[4].

Fig.4. The theory[4] is based on a Gaussian model of a nuclear charge density, a more precise Fermi model can be used in Monte Carlo. If the Gaussian formfactor is used in simulation and numerical integration, results agree within 1%. If the Fermi charge density is used in Monte Carlo the difference between theory and simulation becomes noticeable at large angles.

## ACKNOWLEDGEMENTS

I am grateful to Nikolai Mokhov and Dick Prael for many useful discussions, suggestions and support. This work was supported by the Universities Research Association, Inc., under contract DE-AC02-76CH03000 with U.S. Department of Energy.